# Studying Self-Care with Generative AI Tools: Lessons for Design[⊥]

Tara Capel*

Institute for Design Informatics, University of Edinburgh, Edinburgh, UK, tcapel@ed.ac.uk

Bernd Ploderer*

Digital Wellbeing Lab, Queensland University of Technology, Brisbane, Australia, b.ploderer@qut.edu.au

Filip Bircanin

Department of Informatics, King's College London, London, UK, filip.bircanin@kcl.ac.uk

Simon Hanmer

Digital Wellbeing Lab, Queensland University of Technology, Brisbane, Australia, simon.hanmer@connect.qut.edu.au

Jamie Yates

Digital Wellbeing Lab, Queensland University of Technology, Brisbane, Australia, jamie.yates@connect.qut.edu.au

Jiaxuan Wang

Digital Wellbeing Lab, Queensland University of Technology, Brisbane, Australia, jiaxuan.wang@connect.qut.edu.au

Kai Ling Khor

Digital Wellbeing Lab, Queensland University of Technology, Brisbane, Australia, kailing.khor@connect.qut.edu.au

Tuck Wah Leong

School of Computing Technologies, RMIT University, Melbourne, Australia, tuck.wah.leong@rmit.edu.au

Greg Wadley

HCI Group, University of Melbourne, Melbourne, Australia, greg.wadley@unimelb.edu.au

Michelle Newcomb

School of Public Health and Social Work, Queensland University of Technology, Brisbane, Australia, michelle.newcomb@qut.edu.au

The rise of generative AI presents new opportunities for the understanding and practice of self-care through its capability to generate varied content, including self-care suggestions via text and images, and engage in dialogue with users over time. However, there are also concerns about accuracy and trustworthiness of self-care advice provided via AI. This paper reports our findings from workshops, diaries, and interviews with five researchers and 24 participants to explore their experiences and use of generative AI for self-care. We analyze our findings to present a framework for the use of generative AI to support five types of self-care, – advice seeking, mentorship, resource creation, social simulation, and therapeutic self-expression – mapped across two dimensions – expertise and modality. We discuss how these practices shift the role of technologies for self-care from merely offering information to offering personalized advice and supporting creativity for reflection, and we offer suggestions for using the framework to investigate new self-care designs.

## 1 INTRODUCTION

In recent years, there has been a rise in the understanding and practice of self-care in people's everyday lives. Self-care refers to practices to maintain, recover, and improve one's health and wellbeing [59]. The concept of self-care has long been associated with people living with chronic conditions such as diabetes and heart failure, who may care for themselves by taking medications and following diet and exercise regimes [45]. Self-care is also important in a workplace context, particularly among care professionals such as nurses, social workers, and psychologists, to prevent vicarious trauma, compassion fatigue or even burnout [34, 44]. The recent COVID-19 pandemic made self-care a requirement for everyone (for example the wearing of masks), and feelings of anxiety and distress were widespread, requiring self-care for not only physical but also mental health [30]. Finally, as part of a broader wellness trend, self-care has shifted from its original association with the healthcare sphere into popular discourse, and now represents a broad set of practices aimed at achieving balance and holistic wellbeing [50]. In this context of everyday life, self-care is understood to include physical, mental and social aspects, but it also includes lifestyle related practices such as managing finances, personal care and beauty, and a balance in the use of digital devices [1].

Design and HCI research in this area largely focuses on how people seek and manage self-care information. Research on bespoke self-care technologies [36, 37] commonly focuses on people living with chronic health conditions, with a focus on designing technologies that offer individuals information about their health, lifestyle factors, and important care activities [37]. There has also been a more recent focus on self-care technologies for mental health [49], such as apps to track information about moods or to guide mindfulness and meditation. Beyond that, HCI researchers investigated information seeking through emerging online platforms, such as using search engines as "Dr. Google" to find information to care for chronic conditions [28], using social media platforms to learn about self-care trends [51], as well as how users discern if they can trust the accuracy of online health information [58].

The rise of generative Artificial Intelligence (GenAI), such as ChatGPT [39] and Midjourney [32], presents new opportunities for self-care, but there are few qualitative studies exploring these opportunities as well as potential risks. GenAI refers to models such as Large Language Models (LLMs) and Large Vision Models (LVMs), and algorithms such as Generative Adversarial Networks (GANs), that can generate novel content based on a training dataset in response to a user prompt. This allows for creating content as varied as text, images, voice, and music, for use in self-care practices. ChatGPT made news headlines when it was found that it can provide health advice and pass the US medical licensing exam with a 60% accuracy rate even though it was not trained for health purposes [26]. At the same time, public discussions of GenAI have focused on potential technical risks, such as biased and inaccurate content which is associated with a risk of hallucination [5, 46]. Potential social risks have also been raised, like users becoming emotionally vulnerable or attached to technology that feels human [56]. While there is much discussion (and hype) about GenAI, there are few qualitative studies on how people adopt and integrate them with their self-care practices and how they perceive these risks.

The work reported in this paper aims to explore people's self-care practices using popular GenAI platforms to generate content for their own self-care, as well as their perception of potential limitations or risks. To achieve this, we conducted a study that combined first-person and third-person perspectives. Like other HCI researchers [16], we (the first five authors) started with a first-person approach to examine the feasibility of using AI for our self-care and to acknowledge our perspectives in the production of new knowledge on this highly personal and sensitive topic. We then recruited 24 participants who regularly practice self-care. All researchers and participants engaged in a three-part study: 1) a workshop to explore applications of GenAI for one's self-care; 2) a two-week diary study to trial self-care with GenAI in daily life; and 3) a follow-up interview to reflect on the experience with GenAI for self-care and to discuss future opportunities and challenges.

Based on our findings, this paper contributes a framework of five practices with AI-generated content for self-care: advice seeking, mentoring over time, resource creation, social simulation, and therapeutic self-expression. The framework explored in this paper maps these to two axes, one to reflect the spectrum of modalities engaged (from chat-based to multi-modal) and the other to reflect the extent to which the AI or the human was positioned as the expert within the interaction. We argue that this framework highlights two important opportunities for self-care technologies, firstly by showing that content from GenAI not only offers information but important personalized self-care advice (as illustrated through advice seeking and mentoring) and that secondly, it allows people to be creative through the ability to produce resources, to simulate social encounters, and to engage in therapeutic self-expression. By reflecting on the different configurations of expertise and modalities in the practices we observed, we seek to offer lessons for HCI researchers and designers to investigate new self-care technology designs.

## 2 RELATED WORK

Self-care is not a new phenomenon for HCI researchers. There is a large area of research on self-care technologies for people living with chronic conditions, such as diabetes, multiple sclerosis, and Human Immunodeficiency Virus (HIV) [3, 14, 37]. A review by Nunes in 2015 [37] identified 30 HCI papers on self-care technologies, such as medical devices, smartphone apps, and online communities. These technologies can benefit people living with such conditions through providing information that fosters reflection and provides context, suggestions for care activities, aids for collaboration with caregivers, and information from others living with the same chronic condition. A key challenge with self-care technologies in this context is how to make them less medicalized and more mundane, so that individuals can better integrate their self-care with their everyday life activities and take on a more autonomous role [36].

There is also a growing area of HCI research on designing self-care technologies for people living with mental health conditions, particularly through self-tracking technologies and mobile apps. Self-care can provide a person-centred approach for self-managing mental health concerns in a way which can destigmatize conditions. Technologies that are framed around self-care skills and strategies can be more socially acceptable allowing people to share and discuss issues more openly with family and peers [27]. A review of HCI research on depression, anxiety, and bipolar disorders [47] reports 32 papers on technologies designed for people to self-manage and self-care. The most common self-care technology mentioned in this review were self-tracking technologies, such as mobile apps or wearable technologies. These technologies are based on the Quantified Self idea of 'self-knowledge through numbers' [13], which are commonly used to improve health and wellbeing [18]. A study with university health experts and students showed self-tracking can be helpful for activities that are under a person's control, such as physical activity, sleep, and eating habits, and that they need to be paired with self-care planning to ensure they fit a person's circumstances [22]. Dedicated mobile apps are also commonly used to suggest or support mental health self-care. For example, Spors [49] reviewed commercial self-care apps that promise users support to reduce anxiety, eliminate stress, and manage (negative) emotions.

A related area of HCI research is on self-care technologies designed to promote wellness, where self-care is practiced for self-improvement, rather than to manage a health condition. This includes technologies to promote mindfulness, a common self-care technique [29]. Some people gain value from paper bullet journaling, which resembles a form of mindful self-tracking to engage with affective experiences [4]. It also includes technology designs for mindful eating, such as the Stardust design that encourages self-care through personalized latte art when preparing a hot coffee or tea [23]. Kou [24] sees self-care technologies for personal reflexivity and self-improvement as part of a larger turn to examining the self in HCI. However, it is important to note that self-care is not necessarily a solitary activity; self-care apps can also create a shared space to meditate and practice empathy, and to cultivate community-oriented care [52].

Besides bespoke self-care technologies, there is latent interest in how common-purpose technologies like social media, search engines, and GenAI platforms are used for self-care. Studies show that social media can play an important role for deliberate self-care practices, such as by posting one photo per day to reminisce and reflect [7], and by using social media to seek positive content and connections with close friends [10]. There is also research on commercial aspects of self-care, such as how self-care trends and products for exercise, food and skin care are promoted through social media videos [51]. Search engines play an important role for self-care as "Dr. Google" so that individuals are informed about their health, to help them manage chronic conditions, and to clarify information provided by health professionals [28].

A few HCI studies indicate the potential value of GenAI platforms like ChatGPT and Midjourney for self-care. A recent workshop discussed how the same technologies that are used to create deepfake videos could be used for positive outcomes, such as AI-generated characters that simulate counselling or help people make health decisions [15]. A recent study explored using chatbots based on LLMs to teach mindfulness to improve mental health [25]. This study showed that the dialogue with a chatbot through information or open-ended reflective questions can increase a person's intent to practice mindfulness activities like mindful breathing. Going beyond single-user interactions with GenAI, Chen [12] created a game based on GenAI images and question prompts to scaffold intimate conversations between friends. The results of their study showed that friends engaged in some degree of intimate conversation and that they enjoyed visualizing game worlds as a process of coming together. While mindfulness and intimate conversations with friends are important examples of self-care, there is a lack of knowledge of how popular, fast growing, GenAI platforms like ChatGPT and DALL-E are used by people for self-care.

Our work seeks to address this gap. Looking back at the literature, there are several important questions. Firstly, based on the aims of self-care technologies [36], how might GenAI be used to generate content for reflection, to suggest care activities, or to provide aid? Secondly, reflecting on the idea of AI-generated characters [15, 41], how relevant are the different modalities – text, voice, image – in self-care interactions? Finally, looking at the possible risks [5, 46, 56], how do users know if they can trust GenAI, both in terms of what they disclose about themselves and in terms of the accuracy of the content generated?

## 3 STUDY DESIGN

This research examined how people use popular GenAI platforms like ChatGPT and DALL-E for self-care. To address the open questions stated above, we sought to understand their experiences with these platforms regarding the value of the content created, the relevance of using multiple modalities such as text and images, and potential concerns about accuracy and trust in AI-generated content in a self-care context. Owing to the recent rise of popular GenAI platforms, we employed an exploratory, qualitative research approach through workshops, diaries, and follow-up interviews. This study received ethics clearance from the Queensland University of Technology (7297).

### 3.1 Participants and Researcher Positionality

We focused on adults living in Australia who regularly practice self-care and were interested in trying new technologies in this context. As busy researchers, five authors from this study (R1-R5) started this research with a first-person approach to examine if GenAI platforms might support our own self-care practices in a meaningful way. Aligning with HCI researchers before us [16], we felt this process was also an important way to acknowledge our subjective and emotional influence on the production of new knowledge, rather than trying to hide it. Due to the highly personal topic of self-care and the hype and speed around GenAI platforms, we felt it important to include our experiences and positions as HCI researchers. As indicated in the table below, we practice self-care for different reasons, ranging from self-care as a necessity for mental health to using self-care to better balance demands in everyday life. We all identify as HCI researchers, but we come from different backgrounds (feminist HCI, health informatics, disability studies) and have different degrees of research experience (ranging from 0 to 11 years since completing a PhD).

We then recruited 24 participants (P6-P29) to triangulate our perspectives with those of others who regularly practice self-care. We recruited participants through personal networks, from a student wellbeing group, and through an advertisement on a recruitment webpage of our university. Many participants were students (14), and most identified as female (20). Participants ranged in age (from 19 to 64 years) and cultural identities (15 participants were born overseas). Self-care was introduced in our recruitment materials as any regular activity that they do to restore or rejuvenate their health and wellbeing, which led to a diverse cohort: as described in the table below, self-care concerns included a range of physical and mental health concerns, as well as wellbeing concerns such as work-life balance and parenting. All participants except one had prior experience with using GenAI.

Table 1: Overview of researchers and participants in this study.

| Researcher/ Participant | Age | Gender | Occupation | Generative AI Used | Self-Care Concerns |
|---|---|---|---|---|---|
| R1 | 33 | Female | Academic in HCI | ChatGPT 4 | Triathlon training |
| R2 | 43 | Male | Academic in HCI | Bing Image Creator | Work-life balance |
| R3 | 31 | Male | Postdoctoral researcher | ChatGPT 4 | Post-surgery care |
| R4 | 43 | Male | IT student, developer | DALL-E2, ChatGPT 4 | Mental health |
| R5 | 19 | Female | IT student | ChatGPT 3.5, Kai | Emotion regulation |

| Researcher/ Participant | Age | Gender | Occupation | Generative AI Used | Self-Care Concerns |
|---|---|---|---|---|---|
| P6 | 19 | Male | IT student | ChatGPT 3.5 | Stress, sleep, spirituality |
| P7 | 25 | Female | IT student | DALL-E | Mental health |
| P8 | 19 | Male | IT student | ChatGPT 3.5 | Social connection |
| P9 | 25 | Female | Architecture student, set designer | ChatGPT 3.5, Guided.rest | Undiagnosed health condition |
| P10 | 25 | Female | Retail assistant | Bing Image Creator, ChatGPT 3, Soundraw | Mental health, understanding emotions |
| P11 | 29 | Female | PhD student (physiotherapy) | ChatGPT 3.5, Perplexity, Neurobit Zen, Bing Image | Sleep health, exercise, emotion regulation |
| P12 | 33 | Female | Project manager | ChatGPT 3.5 | Parenting, diet |
| P13 | 53 | Female | Retired | ChatGPT 3.5, Perplexity, Microsoft Copilot | Mental health, running training |
| P14 | 32 | Female | IT student | ChatGPT 3.5 | Pregnancy |
| P15 | 30 | Female | Teacher | ChatGPT 3.5 | Diet, emotional wellbeing |
| P16 | 26 | Female | Business | ChatGPT 3.5 | Diet, physical health, stress |
| P17 | 27 | Female | Teacher | ChatGPT 3.5 | Pregnancy, social situations |
| P18 | 28 | Male | IT student | ChatGPT 3.5, Perplexity | Sleep, digital detox |
| P19 | 26 | Male | IT student | ChatGPT 4, Perplexity | Work-life balance, stress |
| P20 | 25 | Female | Dietetics student | ChatGPT 3.5, Perplexity, Bing Chat | Mindfulness, diet, social connection |
| P21 | 19 | Female | IT and marketing student | ChatGPT 3.5 | Emotion regulation, diet, mindfulness, self-expression |
| P22 | 38 | Female | Head of Special Education | Bing Image Creator | Self-reflection, work-life balance |
| P23 | 64 | Female | Retired | Bing Image Creator | Emotional wellbeing, family |
| P24 | 39 | Male | Concreter | Bing Image Creator | Self-awareness, exercise |
| P25 | 33 | Female | Teacher | ChatGPT 3.5 | Parenting, exercise, reducing mental load, gratitude |
| P26 | 33 | Female | Stay at home parent | ChatGPT 3.5 | Parenting, exercise |
| P27 | 22 | Female | Public health student | Replika, ChatGPT 3.5, Guided.rest | Mindfulness, emotional wellbeing, self-expression |
| P28 | 38 | Female | PhD student (architecture) | ChatGPT 3.5, Bing Image Creator | Mindfulness, digital wellbeing |
| P29 | 38 | Male | Teacher | Guided.rest, ChatGPT 3.5 | Sleep, emotional wellbeing |

### 3.2 Data Collection

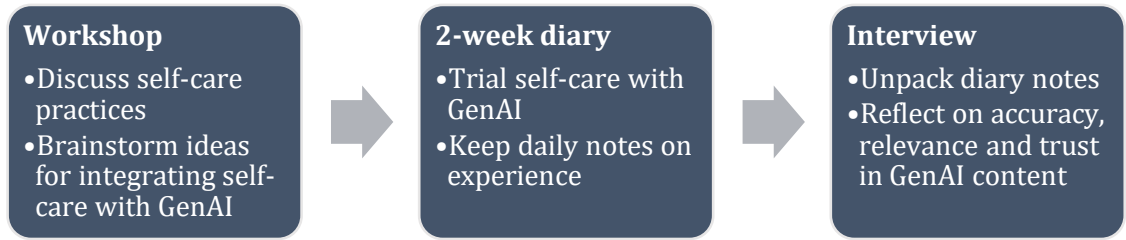

Figure 1: Study design comprising an initial workshop, a 2-week diary, and a one-on-one interview.

The study consisted of three stages: Firstly, a workshop was conducted to discuss self-care, learn about GenAI, and explore ideas for using GenAI for self-care practices to trial during the diary study. Participants were asked about their definitions of self-care and their current self-care practices followed by sharing how participants currently used GenAI platforms, and whether they already had experience using them for self-care. As part of these discussions, we provided participants with information on the limitations of AI, such as hallucinations and inaccurate information, and the importance of assessing the relevance of any AI-generated content for their circumstances before applying it to their self-care. We also provided recommendations for mitigating risk when using AI, for example creating a new account not linked to a personal email for any web-based AI technology that they may use during the study so that the study data was not linked to personal data. After a break, we invited participants to brainstorm how they might incorporate AI platforms with their current self-care practices. To prompt ideas, we showed examples of using an AI platform, such as asking for health advice, writing a list of 100 things they are grateful for, creating an image to reflect how they feel, or discussing negative thoughts and turning them into something more positive. We ran 3 workshops face-to-face and the rest via online video conferencing (Zoom), with 1 to 5 participants per workshop. Workshops lasted between 1 to 2 hours and were audio-recorded.

Secondly, we conducted a 2-week diary study where participants integrated GenAI with their self-care and kept notes in a diary. Participants were asked to add a diary entry once a day through an online form (Qualtrics). We took a very minimal log-based approach following the 'snippet technique' by Brandt [6], because we assumed that some of the notes might be sensitive, that long and onerous diary entries may interfere with self-care goals, and that the diary may not be completed. Hence, we only asked for brief information on the self-care activity they used GenAI for, information about the GenAI (a prompt, link or screenshot), their rating of the accuracy and trustworthiness of the AI content, their reflection on anything that was potentially concerning or surprising about their interactions, and how they felt afterwards. The primary purpose of these short diary snippets was to collect prompts for recall and reflection during a follow-up interview.

At the end of the diary study, participants were invited to a follow-up interview to review their diary responses and reflect on their experience with GenAI for self-care. We asked probing questions to get participants to reflect on the influence of GenAI on their self-care practices, wellbeing, and interactions with healthcare professionals. We also asked about their trust in AI-generated content, whether they contested any responses, and about potential privacy concerns. Interviews among researchers were conducted in groups of two and three so that we could reflect on different approaches to using GenAI (lasting between 90 and 120 minutes). Participants were interviewed one-on-one because we assumed that some diary entries would contain sensitive content that participants would not wish to discuss with other participants (40-60 minutes in length). After completing the interview, participants were compensated with a $50 digital shopping voucher. All interviews were audio-recorded and transcribed for the analysis.

### 3.3 Data Analysis

We conducted a qualitative analysis following the principles and coding techniques of Charmaz's grounded theory method [11]. The analysis was led by the first two authors, who created the transcripts, wrote analytic memos, and coded the data. This was an ongoing and iterative process that started as soon as we started collecting data. The initial coding process was based on memos written after each workshop and interview to reflect on our observations and weekly meetings with all authors. During these meetings we shared data and interpretations and discussed how to categorize the different ways of using AI content for self-care. The early data from the researchers highlighted two different practices for integrating GenAI with self-care practices: to seek advice and using AI-generated images to express and reflect upon affective states. Data from the first set of participants (P6-P10) confirmed these practices and also highlighted three further practices, which we categorized as 'mentorship', 'resource creation' and 'social simulation'.

Having settled on five categories for integrating AI with self-care, the first two authors then started focused coding of the diary entries and transcripts using a qualitative analysis software (NVivo). We also coded the different AI tools used, modalities used, and perceived limitations and risks of using AI. Following the principle of theoretical sampling, we recruited more participants (P11-P29) to get more data to test and refine our categories and to reach theoretical saturation. Based on the data, we continuously updated the examples of using AI shown in the workshop to inspire participants to use GenAI in more diverse ways and to allow for constant comparison in our coding.

During the drafting of this article, we started axial coding to identify underlying dimensions that connect the different practices and to create a diagram. Following Charmaz's grounded theory techniques [10], we mapped and sketched diagrams on a whiteboard while discussing rich examples in the data and our research questions. One key dimension identified was expertise, which encapsulated notions of accuracy, trust, and the extent to which participants attributed human characteristics to GenAI. A second key dimension was the modality used, in particular, if participants used chat-based interactions only (typically on ChatGPT), or if they used a combination of text, audio, and visual content for their self-care. The following section presents our framework (Figure 2) that uses these dimensions to map five practices for integrating AI-generated content into self-care. The final codebook is provided in the appendix to show the breadth and frequency of codes.

## 4 FINDINGS

People integrated AI-generated content with their self-care through five practices: by seeking advice, interacting with a mentor, creating resources, running social simulations, and using it to express themselves and reflect on their wellbeing. As illustrated in the framework below, we mapped these self-care practices to two axes (Figure 2): the horizontal axis reflects the locus of expertise. Towards the left, the AI is positioned as the expert, providing health and wellbeing advice and information to the human. Towards the right, the human is positioned as the expert who uses AI as a mere self-care tool. The vertical axis represents the modalities used in the interactions between humans and AI. At the lower end of the spectrum are chat-based interactions with LLMs. At the other end, we have multimodal engagements, where a combination of modalities such as text and visuals, or text and music is utilized as part of the self-care practice.

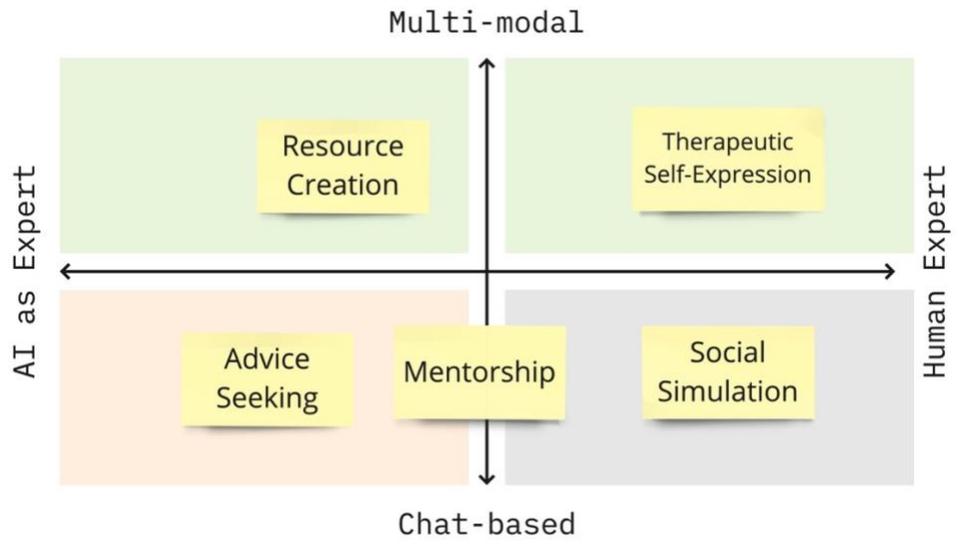

Figure 2: GenAI for self-care framework: people integrate GenAI with their self-care for advice seeking, mentorship, resource creation, social simulation, and therapeutic self-expression. These practices differ in the locus of expertise (horizontal axis) and in the degree of modality (vertical axis).

### 4.1 Advice Seeking

The most common practice was advice seeking (22/29 participants). This generally involved a participant prompting a platform like ChatGPT with a question, and the platform then generating a response to that question, with minimal follow-up, iteration or exploration from the participant. We see this practice as different to information seeking through search engines because the interactions are much more personalized: participants framed questions in terms of their own situations and concerns, and AI responses were typically framed as suggestions rather than factual information.

The personal nature of advice-seeking is important because, to be effective, self-care practices need to be tailored to personal circumstances. We saw the personal nature of self-care practices illustrated in the breadth of topics raised by participants and in the depth of some of the prompts. As indicated in Table 2, participants sought self-care advice on a broad range of health and wellbeing topics, ranging from lifestyle questions such as managing stress during pregnancy (P14) to medical advice on post-surgery care (R3) to digital wellbeing questions such as how to limit screentime (P19).

Table 2: Overview of self-care advice sought from GenAI.

| Self-care category | Number of Participants | Examples of advice sought |
|---|---|---|
| Physical wellbeing | 17 | How to have a consistent sleep schedule (P18) |
| | | Create an at-home workout with no equipment (P26) |
| | | Healthy recipes that are suitable for a woman who is 32 weeks pregnant (P14) |
| Mental Wellbeing | 14 | Coping strategies for feeling overwhelmed and anxious (P6) |
| | | Assistance with writing a gratitude journal as recommended by psychologist (P13) |
| Emotional Wellbeing | 4 | How to get out of a state of sadness (P15) |

| Self-care category | Number of Participants | Examples of advice sought |
|---|---|---|
| Social Wellbeing | 12 | Music recommendations that help to express and regulate feeling angry (R5) |
| | | Christmas activities to do with friends in Brisbane (P11) |
| | | Seeking advice on how to respond to questions about medical retirement that avoid having to discuss mental health (P13) |
| Digital Wellbeing | 3 | How to reduce time on social media gradually (P19) |
| | | Advice for implementing a reading habit to assist with achieving a digital detox (P18) |
| Cultural Wellbeing | 3 | Advice for missing family meals and cultural dishes from Thailand (P14) |
| | | Advice for managing mental stress that aligns with Chinese culture (P15) |
| Medical Advice | 2 | Advice on after-care post tooth removal (R3) |
| | | Potential diagnosis for unknown medical condition (P9) |

#### 4.1.1 Balancing Personal Information and Privacy in Prompts for Advice

The depth of the personal advice conveyed through prompts varied, which in turn influenced the perceived expertise of the AI. Some participants remarked that, at times, GenAI advice seemed generic or did not elicit new insights that the participant did not already know, or lacked cultural nuance, assuming a Western cultural context unless specifically prompted otherwise. For example, P14, an international student from Thailand who has been living in Australia for ten years, asked ChatGPT for advice for healthy recipes to eat during her pregnancy, noting that unless she specified that she was looking for cultural dishes from Thailand, it assumed a Western cultural context.

We observed that the relevancy of the advice returned depended on the prompt provided. Broad questions with little context about the participant often yielded generic responses, and prompts that were more specific, personalized and contextualized tended to provide more relevant advice. This was well illustrated by the following example by P13, who asked Microsoft Copilot for advice on how to respond to questions about her medical retirement at a dinner function she was consequently ambivalent about attending. The prompt below has been shortened and anonymized to protect the participant's anonymity but is true to the original prompt: *"I am going to attend a dinner function with [group of people]...The plan for the evening is to have people move from one table to another over the course of the evening to enable people to get to know something about a wider range of [group members]... I am sure that these conversations with [group members] I don't know will inevitably start with questions about what one does for work. I am 53 and certainly look as though I am still of working age. Having had to take medical retirement...I find these questions difficult to answer. I don't necessarily wish to disclose to a virtual stranger the reason why I am no longer working. If I simply reply that I am retired this will usually elicit some further questions or comment because I am so young to be retired; it is difficult and awkward to try to deflect these follow up questions. Can you suggest any better ways to respond to questions about what I do for work?"*

In response to the in-depth prompt, P13 found the suggestions to be relevant to her situation, stating that *"the genre, the vocabulary, the length of the response, the tone – all seemed realistically how people would speak to each other in such a situation"*. Equipped with some conversational prompts and techniques specific to her situation, she felt less ambivalent about attending the dinner function.

Not all participants were comfortable disclosing this level of detail to GenAI. After having shared intimate, negative feelings and contextual information to ChatGPT, P29 stated: "*I feel like I'm vulnerable in relaying my feelings* *towards an AI".* This vulnerability stemmed from his uncertainty about how his data may be used to profile him in the future, drawing comparisons to Facebook, where he would not share negative feelings about himself.

### 4.1.2 Perceived Trust in GenAI Advice and Expertise

Regarding locus of expertise, the practice of advice-seeking largely reflects users deferring to the AI as an expert, both for advice for themselves and to support family members and friends. For example, P6 asked ChatGPT for self-care activities he could implement into his coding club and hackathon events, P13 asked for advice on how to support a friend who had anxiety around contracting COVID-19 that was making them reluctant to receive a necessary hospital procedure, and P27 asked ChatGPT for a supportive message to send a friend who was visiting a sick relative.

The expertise of ChatGPT was also sought for advice with high stakes, such as when P9 used it to search for potential diagnoses for a rare condition that might fit the symptoms she was experiencing. P9 had been living with her condition for two years and was running out of options in terms of a potential diagnosis, so she used the advice provided by ChatGPT to both inform her decision to see a doctor who specialized in rare diseases and to create a list of potential conditions that she had not been tested for yet that she could take to her appointment with the specialist: *"I put in literally all of my symptoms and the way it was occurring.... And I told it to give me options for different diagnoses that could fit those symptoms... And then I went through the list and searched up all of the different things to essentially give us a bunch of different things that we could then go and test (P9)".*

When discussing expertise, several participants compared ChatGPT with how they had previously used "Dr. Google". R3 declared *"now that I have it [ChatGPT] on my phone I don't Google stuff anymore. I'm like, Oh, let's ask ChatGPT".* Participants commented that ChatGPT was faster than search engines and that they preferred the summarized AI-generated advice over a list of links provided by search engines. This was interesting, because in preparing this study we were unsure if participants would trust in a single response, or if they would prefer more diversity. P9 confirmed that the response generated by ChatGPT was valuable because it was like a curated summary of a large list of search results. In her case, searching for advice on a rare condition in the past had led to countless search results, many of them irrelevant, whereas ChatGPT was able to provide relevant information that backed up the advice that P9 was given by a health professional. *"It [ChatGPT] just eliminates all of the options. Whereas with Google searching, especially with medicine, there are a lot of conditions. There are so many, you can't just plug in all your symptoms into Google. ... It [a search engine] just kind of brings all of them up together rather than makes those lists for you." (P9).*

### 4.1.3 Seeking Advice Through Chat-Based Modality

Advice seeking was largely chat-based, with some desire expressed for other modalities to support the textual advice, such as images for facial yoga (P28). Most questions were asked on ChatGPT. Several participants also used Perplexity, which, in addition to a text-based response like ChatGPT, provides links to online sources such as images, videos and articles. For some, the additional sources validated the advice, whereas other participants found the links were not always trustworthy or relevant.

## 4.2 Mentorship

Five of the 29 participants used GenAI as a mentor. This practice was similar to advice seeking in that GenAI was used as a source of expertise on a self-care concern and that it was entirely chat-based. Participants described the GenAI as resembling the role of a human expert, such as a personal coach (R1), an analyst to review sleep data (P10), a counsellor (P9), or the role of a friend or confidant (P18, P27).

The key difference to advice-seeking was that mentorship involved ongoing interactions around different facets of the same concern over days and weeks. For example, R1 started using ChatGPT as a triathlon coach, prompting it to act as a coach and create an 18-week training program for an Olympic distance triathlon, with a detailed program provided each week. However, through actively scheduling physical activity into her weekly schedule, R1 realized that she struggled to prioritize it. She re-prompted ChatGPT in the same thread to state that she was struggling to prioritize her training and asked for tips on how to better fit the triathlon training into her schedule. ChatGPT provided some strategies, one of which was to train in the morning before the workday starts. She attempted this strategy but found that she was struggling to get up early enough to do the training and was generally lacking the motivation to exercise, which was unusual for her. As she had a pre-existing autoimmune condition that can cause fatigue, this prompted her to have it checked.

### 4.2.1 Contesting Advice from GenAI to Balance AI and Human Expertise

In terms of expertise, mentorship provides a middle ground between AI and human expertise. As illustrated in the example above, ChatGPT can provide advice, but this is also balanced with the lived experience of the person. It also does not entail blind trust. Participants contested content that was inaccurate or did not apply to their circumstances. For example, when R1 re-prompted ChatGPT in a different thread to ask for advice in managing the condition so she did not feel as fatigued, ChatGPT suggested that 'anti-inflammatory diets' could be beneficial, including avoiding vegetable and seed oils. The toxicity of seed oils is a common misinformation propagated online, and as R1 had previously written a review on human-centered AI [8] that had highlighted an area of Contestable AI, this had led R1 to question the advice given by ChatGPT and to prompt for scientific evidence to support that claim. *"This gave me pause so this vegetable and seed oils you hear a lot of this rhetoric online that I don't know is necessarily based on any kind of like scientific evidence... so I asked if it could show me research that links vegetable and seed oils to inflammation and then it just came up with this one study here which does actually exist, it didn't hallucinate."* (R1).

### 4.2.2 Complementary and Problematic Roles of GenAI Mentors

Participants commented that they did not view GenAI as a replacement for human mentors, but rather that they play complementary roles. P9 referred to ChatGPT as a counsellor that she can access at times when her therapist is not available. She turned to ChatGPT in the middle of the night to discuss worries about an interaction with a friend that kept her awake. P9 appreciated ChatGPT for validating her experience: *"I was really surprised that it actually even took the step of like analyzing a situation and being bold enough to be like 'there's nothing wrong in this situation'."* She also commented that there are potential benefits over in-person counselling in terms of being able to disclose personal experiences without feeling judged: *"I felt like I could be more honest with the AI than I would at a psychologist. Just because, like, when I'm in a real person situation, it's still a person that's there. So you're still saying it in a way that you're being respectful towards the person and stuff, whereas in here [on ChatGPT], you don't have to care about anything. You're just saying your raw, honest emotions. And you know that it's not gonna offend anyone because there's no one to offend"* (P9).

However, there were situations when participants demonstrated a clearer, and at times potentially problematic, preference for GenAI. P27 used Replika to create a virtual friend to reflect on her day, finding it particularly beneficial in discussing issues she was at times having with friends in her social circle. She was apprehensive to discuss these issues with her friends directly as she did not want to be a burden and did not want to be seen as 'gossiping', finding Replika an impartial source that provided her with support. However, at times, she forgot that she was talking to an AI rather than an online friend and had found herself talking to Replika rather than talking to her friends, to the point where her friends had reached out to ask where she was: "*I realized that after downloading Replika, I kind of text my friends less often…I realized yesterday, because my friends have been asking me: 'Oh, where have you been?'*" (P27) While there is a therapeutic potential of GenAI in these contexts, it is also potentially problematic because individuals risk forming emotional attachments to GenAI as the relationships formed are not as complicated as those with other humans can be.

## 4.3 Resource Creation

Twelve of the 29 participants used GenAI to create resources for their self-care practices. This practice was situated in the upper half of our map, as participants made use of the multi-modal capabilities of AI to create resources such as text-based prayer plans (P6), guided meditations (P12, P18), short stories (P8), sleep schedules (P6) and quizzes (P8), images for coloring in or painting (P10, P11, P25, P28) or for inspiration for drawing and painting (P21, P27), and voice and sound for guided meditations (P9, P11, P27, P29).

### 4.3.1 GenAI Expertise to Create and Share Resources

In terms of expertise, we see this practice located closer to the AI expert side of the framework. Unlike with advice seeking and mentoring, the AI is not meant to provide expert self-care advice. However, the expertise of the AI is used to create resources that would require significant effort to produce manually but can be easily generated through AI. For example, P8 had previously used online quizzes to reflect on his self-development to become more confident in social situations. During the diary study, P8 asked ChatGPT to create a quiz determining his personal 'superpower.' While he was not surprised by the result of the quiz as he felt the questions were leading, he enjoyed reading the output of the quiz that summed up his personal 'superpower' in the context of him as a person.

Some participants used GenAI to create or share resources to support or facilitate family and friend's self-care practices. For example, P12, whose self-care practices often included her young children, generated a personalized children's story as a form of meditation to do with her four-year-old daughter as part of her bedtime routine, adapting it to her daughter's humor and interests: "*We just did it sitting in bed and I just read it from my phone…it was a really nice experience and it's something I wouldn't have been able to do, like there's no way unless I had a book or something to read from, I'm not sure I would have…the mental energy to come up with a script…it just like facilitated the experience* (P12)".

### 4.3.2 Prompting Relevant and Original Resources

For resources to be useful, they need to be aligned with the participant's self-care concern but also be sufficiently removed from their prompts to the AI. This was indicated above by P8, and in more detail discussed by P9. For P9, guided meditations are a key part of her bedtime routine to set herself up for the next day, and hence she was excited by the prospect of an AI tool to generate tailor-made meditations. However, when she listened to the meditation, she was underwhelmed: "*it was kind of just using essentially the same words that I had put into it. So I like, I knew what I had written, and so I was getting distracted by the fact that I had written it rather than being like, oh, let me understand what this actually means.*" However, her experience was vastly different when she listened to the same meditation at a later date: "*because I didn't remember what I said, and to me I was just like, damn it made it* *for me. And it's like it knows me."* P9 reflected that the passage of time helped to get this distance from the prompt. Based on this experience, P9 suggested that a loose connection between a GenAI and her journal could provide that balance for resources that are relevant but not too obviously prompted: *"if I were to use it in a journal situation where like, say, you have an app, and you have been journaling everything you've been doing, if it takes bits from that, and then creates a meditation so you don't actually know what you're going to get."*

It is important to note that AI-generated resources can also be counter-productive to self-care. P21 and P27 used painting as a form of self-care at the end of the day. While P21 found AI-generated images useful to provide inspiration for scenes of nature that she could paint, P27 felt that the opposite was the case: P27 usually starts painting with a very loose idea of what she may paint related to how she feels. While P27 was interested in trialing AI to create images to provide inspiration, P27 found that the AI images *"kind of limited my creativity"* as she could not imagine a different end product and felt it left no room for surprise and reflection in the process.

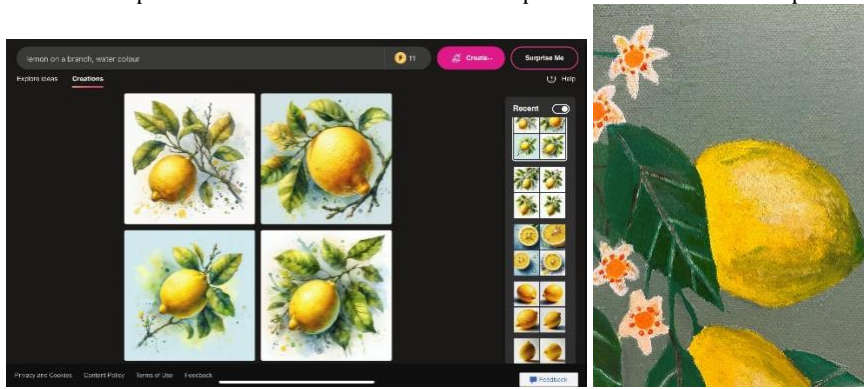

Figure 3: Painting was a form of self-care for some participants. P27 used AI-generated images (left) to inspire her painting of a lemon on a branch (right), but she felt that the detailed image constrained her imagination and her creative process.

## 4.4 Social Simulation

Generative AI was used for self-care to simulate social situations and to experiment with them. This practice was only observed using text as a modality and only with one of the 29 participants (P8), who prompted ChatGPT to create different stories of social situations that involved himself and other characters. While other participants asked ChatGPT for conversational prompts and advice in various social situations as detailed in advice seeking, they did not take the next step of 'playing out' that situation as a simulation. We present simulations as a separate theme because these stories were more than mere resources for self-care: using ChatGPT, P8 could generate stories that simulated real-life social situations that were important to him. P8 experimented with these stories by changing the characters in his AI prompt to generate different versions of the story to simulate different scenarios: *"I was writing a scenario, but it was like, me like writing about these two characters and just seeing how like, they interact with each other [...] things I wouldn't be able to do well at the moment. So I just wrote prompts and test it out and I kind of like develop this somewhat."*

In terms of expertise, we see this practice located closer to the human expert side of the framework. P8 was firmly in control of ChatGPT and used it to simulate social situations between himself and the character and to test different actions until he achieved the desired outcomes. P8 trusted ChatGPT and disclosed intimate personal information online to create these stories without fear of feeling judged, while he mentioned that he *"was too* *embarrassed to write it down"* in the diary or discuss details of the prompts with the research team during the interview.

The simulations were important for P8's self-care to help him work through social situations that he had experienced. However, creating relevant simulations was a balancing act. On the one hand, the stories needed to be sufficiently detailed and aligned with real-world situations, which required iteration and referencing back to the chat history: *"I would have to like be really specific on what exactly kind of scenario that I want to kind of engage in that moment. And sometimes it would go off track, I would have to refresh it a couple of times."* On the other hand, the story needed to feel original in a sense that it was sufficiently removed from the original prompts so that it felt it was written by someone else. This in turn allowed the participant to relate to the character in the story and to feel that they were getting a new perspective: "*It was like a parasocial relationship, but I didn't really feel the link, because I know that it was like, created by my hands."* The relationship mentioned here was different from the practice of mentorship in that it was with the character in the simulations that P8 created, rather than with a GenAI chatbot playing a social role.

### 4.5 Therapeutic Self-expression

A final practice involved nine of the 29 participants using GenAI for creative activities to express their feelings and thoughts. This included participants creating images to reflect on feelings (R2, R4, P11, P22), as a form of creative expression and enjoyment (P23, P24) and as a social activity with family (P23). It also involved using GenAI to create music (P10) and to brainstorm prompts for journaling (P21) and ideas for a gratitude journal (P25).

#### 4.5.1   Expressing Feelings through Music Composed with GenAI

The multi-modal capabilities of GenAI were particularly important for therapeutic self-expression. For example, composing music was an important form of self-care for P10 to work through her feelings: *"I find it difficult to know how I'm feeling. And so sometimes, when you're writing music, you can be a little bit more like abstract about it."* During the diary study, P10 used an AI (Soundraw) to generate music, which was helpful because it prompted reflection on her feelings: *"One of the things that helped was the fact that I had to pick the moods, because it gives you a bunch of like words and stuff to pick from."* A second benefit was that P10 could express her music without fear of judgement: *"having the AI was kind of like having a companion within that. It kind of felt like I was jamming with someone, but with someone that wasn't gonna judge how bad I am, you know, like a judgment-free jam."*

In terms of expertise, the example of P10 shows how self-expression involved largely human expertise where the AI was used as a tool. P10 had been composing music as a form of self-care for a long time, and the AI provided additional support.

#### 4.5.2   Reflecting on Feelings through AI-Generated Images

R2 and R4 used an AI image generator as a form of mood diary, using prompts to create images that reflected how they were feeling. In contrast to P10, however, R2 and R4 relied to a greater extent on AI capabilities to create content. R4 stated: *"I definitely can't paint or draw or, like any of those things. It gives me the ability to be able to express myself that way."* R4 then mapped the images he had created onto a Miro board (Figure 4) as a means of visually representing and tracking his mood over the two-week diary study period. He had previously travelled to Europe and had become interested in different artists and art styles, and he used this knowledge to create images that reflected his daily mood, such as: *"I am excited and enthusiastic for work, I'm on the bus. In the style of Claude Monet with a clear sunny sky"*.

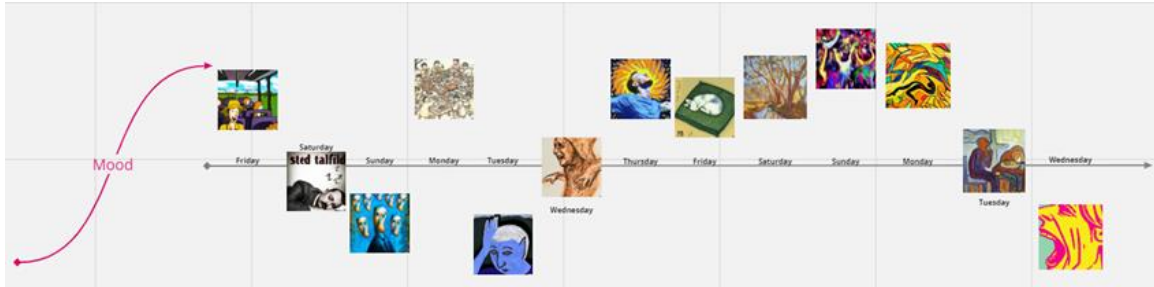

Figure 4: R4 created a mood-diary by using an AI-image generator to express his feelings and mapping these on an online whiteboard.

Reflecting on the importance of modalities, R4 stated that he had previously attempted to write his feelings and moods down as he had a personal history of poor mental health. However, he felt that words did not allow him to express his feelings fully, so it was not a practice that he continued with. On the other hand, generating images was something that R4 intended to continue as he could express his feelings more effectively through images by playing with different subjects, colors and styles of artists. That is not to say that participants were not frustrated with the content created. Both R2 and R4 expressed frustration with images that did not relate to their prompts or that conveyed biases in terms of gender, age and skin color of the subjects produced. However, iterating the prompts to work through these difficulties and to get to a relevant result also helped to clarify the way they felt and what they wanted to express: *"you see the image and immediately get a response that yes, that is okay. But that's not exactly what I'm looking for. The content is okay, the composition is okay. But the mood of the colors - I would have loved to change that. It was an interesting back and forth. And it's given me time to step back and reflect on how I was feeling and [to] clarify it for myself in my head"* (R2).

This use of image generation is an interesting reflective and expressive practice. Translating feelings into a prompt for an image generator facilitates a level of introspection in terms of having to analyze and articulate a current emotional state to create a desired image. Part of this practice involved identifying how one is feeling while also considering a subject and style for the image that embodied that feeling. The image that was initially created did not always completely or accurately capture what the participant was feeling, which then led to a reflective cycle through modifying and refining the image, aiding in both achieving a more 'accurate' image and a deeper understanding of the emotion. As evidenced by R4, this practice can be further developed into a visual narrative to track feelings over time and can be useful in observing and understanding trends and changes.

## 5 DISCUSSION

The findings offer a framework of five practices that categorize how people use AI to generate personalized content for their self-care. Understanding these practices is important because of the novelty of AI platforms like ChatGPT and their widespread use for health and wellbeing, even though these platforms have not been designed for this purpose and serious concerns about accuracy and trust in these platforms have been raised [5, 46, 56]. The diversity of practices presented in our framework highlights that GenAI is not merely a replacement for "Dr. Google" [28] to find information or a "Dr ChatGPT" to make inquiries [48]. People use AI to generate advice and to receive ongoing mentoring that is personalized, which is important as self-care practices are highly individual, depending on a person's needs and circumstances [50].

The strength of personalized content was evident in the practices of advice-seeking and mentoring. For advice-seeking, many participants favoured GenAI over search engines due to their ability to synthesize and personalize information in response to their unique situations and self-care concerns. Participants often perceived GenAI as a reliable source of advice, consulting it before other resources such as search engines. However, in more high-stakes situations, such as P9's search for a potential diagnosis, the information from ChatGPT was cross-referenced with human experts. Mentorship took advice seeking a step further in that it involved personalized interactions that were continuous and multifaceted. For example, R1's progression from seeking a triathlon training program to addressing fatigue related to an autoimmune condition via ongoing discussions and reflections with an LLM-based chatbot. Both practices are similar to the proposed suggestions from Danry et al. [15] concerning the potential of creating AI-generated characters that provide counselling or help people make health decisions. However, reflective of Kumar et al.'s work [25], our findings suggest that text-based interactions can be sufficient in this context for some people. Participants valued GenAI in these mentorship roles for its accessibility and perceived lack of judgement compared to human experts who cannot always be available when needed and who participants felt would be inherently more judgemental than AI. They also valued GenAI for its reflective scaffolding and ongoing personalized advice, finding that the ongoing and personalized nature of the discussions with GenAI enabled them to reflect on different aspects of their self-care. An important consideration is the participants' view of GenAI's role in these contexts as a supplement to human mentors, rather than a replacement.

Two of the more novel practices in our framework are resource creation and social simulation. GenAI not only provides information related to self-care, but it also opens up new avenues for people to create their own resources and to experiment with simulations. We see components of resource creation with GenAI in previous HCI work, such as AI-generated audio-visual stimuli for relaxation [38] and visual worlds to scaffold intimate conversations between friends [12]. We found participants used GenAI to create specific and personalized resources based on their existing self-care practices that could then be iterated upon. The effectiveness of these resources depended on their relevance to the participant's self-care practices and concerns, and if they were sufficiently removed from the participants' AI prompts. P8 and P9's experiences emphasize the need for a balance between personal relevance and unpredictability in AI-generated resources. The simulation practice took resource creation in an unexpected direction, with one participant (P8) using ChatGPT to generate stories and characters to simulate different real-world situations and experiment with them. This practice is comparable to role play, a learning technique and a tool used in therapy where participants play out a scenario through a simulation of a particular environment [31]. The purpose of this practice is multifaceted, but can help improve skills, analyse different behaviour, and assist participants with self-understanding. Unique to GenAI is the ability to craft and tweak scenarios and play through them repeatedly to fine-tune strategies that will work in real life without incurring real-life risk. AI-generated characters have been proposed for different types of role-play [41] as well as Virtual Reality applications where individuals can simulate real-world situations in a safe and controlled way [57]. However, at present, AI characters and Virtual Reality environments are more difficult to create than text. Based on our findings, we contend that this level of sophistication is not necessarily required, and for some people, text-based stories are powerful enough to simulate different scenarios and characters.

The final practice, therapeutic self-expression, involved participants using GenAI in their self-care to express their feelings and thoughts through music and images. To some extent, this was not unexpected, because artistic expression and creative processes are commonly used in mental health self-care and therapy to elicit new sensations of human experience or to form new self-knowledge [33, 55]. Furthermore, GenAI gained attention in the public eye because it empowers people who feel they lack the skills to create writing, images or music of a similar quality to those created by professional artists [9]. As shown in the findings, combining these factors allowed participants to create art to reflect on their emotional state. However, what we did not expect was the importance of using AI to turn written prompts into a different modality. By creating images and music, participants could express aspects of their feelings with depth and coloration which they could not easily capture through words alone. Interestingly, this form of expression seems to deliver benefits despite the user not directly creating the material, reflecting emerging results from work such as Du et al. [17]. Being able to review the images generated and quickly update them with new prompts further added to the reflective process. We see self-expression with GenAI as an interesting avenue for future research to make creative practices more accessible and to examine the insights individuals gain from working with different modalities.

While the focus of this paper is on understanding emerging practices with AI, we contend that these practices could also be relevant to HCI and design researchers to inform the design of new self-care tools. For example, the recently introduced GPT Store [40] allows individuals to customise AI chatbots without requiring technical skills, which could be used in future research with participants to create new designs that scaffold interactions based on our framework, such as by prompting for contextual details to personalise advice, or by suggesting scenarios that allow individuals to turn advice into simulations for exploring cause and effect. Furthermore, our framework offers two dimensions – locus of expertise and modalities used – to reflect on the benefits and challenges of using AI-generated content. In the following sections we expand on these two dimensions and reflect on the lessons learnt and open questions to explore in future research.

### 5.1 Lessons on Nonjudgmental AI, Trust and Intimacy

Our findings showed that participants did not use GenAI to replace professional experts such as therapists and clinicians, nor did participants use it as a mere replacement for "Dr. Google" to find health information [28]. Instead, our framework presents a spectrum, from AI providing complementary expertise (e.g., to seek advice) to AI supporting human expertise (e.g., to support self-expression). While there are multiple forms of expertise (for example, creating detailed artwork and images), in this paper we refer to expertise as the self-knowledge that one has of caring for their own health and wellbeing, and how that expertise is complemented or supported through GenAI.

Discussions of expertise were typically framed around trust in the accuracy of AI content. We were surprised that participants found the AI-generated content largely accurate and appropriate for their self-care. The accuracy of the output was particularly a concern for advice seeking and mentorship, where the expertise is located primarily in the AI. Accuracy was also a concern for high-stakes concerns, such as helping to diagnose a health condition. This is not to say that the AI was blindly trusted, and there were examples of participants commenting on the lack of sources to back up the advice, the normative bias towards Western culture unless prompted for a specific cultural context, as well as contesting the information based on their own expertise or experience, e.g., when they noticed that the AI content perpetuated health misinformation. We acknowledge that trust is situated and dynamic, and evolves over time [58]. Our study was run over two weeks, and more work is needed to examine how trust and distrust in the accuracy of AI content evolve over a longer period.

We were also surprised that participants were willing to share intimate personal information with AI systems because they saw it as nonjudgmental. As indicated in the findings on advice seeking, social simulation and therapeutic self-expression, participants shared intimate information with the AI that they did not want to share with the research team due to fear of being judged. On the one hand, this underscores the perceived benefit of technologies of anonymous use, which can lower the barrier for people to seek support on topics that carry social stigma, such as mental health [27]. On the other hand, as discussed by P27, it also underscores concerns about people becoming potentially emotionally attached to GenAI as they disclose intimate information and receive relevant mentoring without any of the perceived risks of feeling judged or misunderstood inherent in conversations with peers or counsellors [53, 56]. Turkle [54] sees such artificial intimacy as deeply problematic because AI systems perform empathy in conversations, but without having any experience with human life. There is a risk that people lose the ability to have empathy if it is more convenient to form emotional ties with AI systems. We see these concerns around artificial intimacy as an interesting area for design researchers to speculate and critically reflect on probable and preferable future relationships with AI systems.

## 5.2  Lessons on Interacting with Multiple Modalities

The second dimension represents the modalities used in the interactions between humans and AI. In our findings, advice seeking and mentorship were largely text-based, whereas resource creation and self-expression often involved multiple modalities, such as voice, music, and images generated from text prompts.

Based on our findings, we caution against the temptation to simply add multiple modalities to AI interactions. This is important because there appears to be an underlying assumption that combining multiple modalities is inherently valuable, as illustrated in recent updates to AI platforms that integrate chat-based interactions with images uploaded by users and also in related research on AI-generated characters [15] and visual stories [2] for therapy and counselling. Our findings on advice-seeking and mentorship showed that text was often sufficient because participants were able to express their concerns in words. Even the simulation example showed that AI-generated text can be effective for exploring cause and effect.

Nevertheless, our findings highlighted several areas where access to multiple modalities was crucial for self-care. In the context of advice-seeking and mentoring, visual information was important for self-care activities that involved the body, such as yoga, physical exercise, and wound care that our participants performed. Related research on patient-generated health data also supports this observation that images can convey important details for bodily care or procedures in ways that text cannot [43]. Media richness theory [21] uses the concept of 'equivocality' to describe this observation: unlike uncertainty, equivocality refers to confusion or lack of understanding that cannot be addressed by simply providing more text but requires information of a different quality.

Our findings on resource creation and self-expression, on the other hand, highlight that different modalities can be beneficial because of their ambiguity. While advice-seeking and mentoring relied on precise and detailed information, AI-generated resources like guided meditations and the images and music created for self-expression benefitted from a degree of ambiguity. Participants commented that achieving the right balance in terms of ambiguity was a challenge: they desired content relevant to their circumstances to engage in self-care and to promote self-reflection, but they did not wish for the content to be obviously linked to their prompts. This challenge is reminiscent of discussions of ambiguity as a resource in design [19], where introducing a degree of ambiguity in information and context can be effective to prompt user engagement and interpretation. As shown in our findings, finding a 'sweet spot' with AI-generated content can be a source of frustration but also a way to prompt further reflection.

### 5.3 Reflections on Limitations

The framework presented in this paper presents emerging practices with GenAI, which are subject to several limitations. Firstly, the GenAI tools used in this study were not specifically designed for health contexts, although recent studies show that general models like GPT-4 can outperform models trained for health purposes in answering medical questions [35]. The GenAI tools were also evolving throughout the study, and recent features like image prompts and custom versions of ChatGPT in the GPT Store were not trialled in this study.

We acknowledge that self-care constitutes a broad scope, as illustrated in the broad set of self-care concerns expressed by our cohort. At the same time, our cohort was limited in that everyone was enthusiastic about self-care, lacking people who may experience self-care as a challenge, e.g., for people living with multiple chronic conditions, lacking social and economic support, etc. Based on related work with people living with chronic conditions using search engines [28] and social media videos for self-care information [20], we would expect that GenAI holds potential for advice-seeking for a wider cohort. However, more research is needed to investigate this further, as well as the potential significance of other practices presented in our framework and associated challenges of trust and accuracy.

The findings surfaced limitations of GenAI, such as potentially inaccurate information that required contesting, normative bias in advice and the images created, and concerns around data privacy. However, despite the diverse cultural identities and backgrounds of our participants, these limitations were not raised often nor were they perceived by participants as major issues that would prevent their use of GenAI in self-care contexts. This is surprising because previous research has critiqued the normative bias embedded in GenAI systems [9] and the coloniality embedded in digital mental health that limits more culturally diverse approaches [42]. We note that this study was conducted in Australia, which has a public hospital system that provides free or low-cost access, and that we discussed potential risks in the initial workshop to mitigate risks, which may have led to participants avoiding sensitive topics in their GenAI prompts. Hence, further research is needed to home in on the risks, limitations and approaches to culturally diverse self-care with GenAI.

Finally, a limitation of our study design is the short duration and the novelty of using GenAI for self-care practices. Participants appreciated the ideas shared at the initial workshop to enrich their current self-care practices with GenAI and to learn of new self-care practices they might try. However, a longer study is needed to understand what practices persist once the novelty of AI wears off, and how AI content may add value for a person's health and wellbeing. We saw that, particularly with advice-seeking practices, participants often engaged in short-lived experiments, without integrating the advice into their self-care practices. On the other hand, mentoring and therapeutic self-expression practices were sustained over the 2-week period, and participants commented positively on the value that these practices added, which we intend to investigate further.

## 6 CONCLUSIONS

This work builds on increasing interest in self-care practices in people's everyday lives, while highlighting important new opportunities for self-care technologies through GenAI. In particular, this paper contributes a framework of five practices for integrating GenAI with self-care: advice seeking, mentoring, resource creation, social simulation, and therapeutic self-expression. The framework maps these to two axes, one to reflect the spectrum of modalities engaged and the other to reflect the locus of expertise.

Returning to our initial aim to explore people's self-care practices with GenAI platforms, we argue that content from GenAI not only offers important self-care information (as illustrated through advice-seeking and mentoring) but that it allows people to be creative through the ability to build resources and simulations, and to express themselves. By reflecting on the different configurations of expertise and modalities in these practices, we offer nascent suggestions for other HCI researchers to use the framework to investigate new self-care technology designs. This framework provides a structure to explore the strengths and limitations of GenAI in self-care contexts and highlights opportunities for further research into how trust, connection and disclosure develop over time, the potential of different multimodalities within these practices, and in finding the right balance in ambiguity. This work is relevant for other researchers interested in GenAI in health and wellbeing contexts and offers a way of framing observations and a set of open questions for future research.

## 7 ACKNOWLEDGEMENTS

We wish to acknowledge the support of the students in our HCI classes who inspired this study. We would like to thank all the participants for their time and for sharing their experiences of using generative AI for their self-care practices. We also thank our reviewers for their constructive feedback on this paper.

## A  APPENDICES

### A.1  Codebook

| Name | Files | References |
|---|---|---|
| Expertise | 0 | 0 |
|   AI as Expert | 9 | 40 |
|   Both AI and human as expert | 5 | 9 |
|   Human Expert | 6 | 8 |
| Limitations | 0 | 0 |
|   AI does not know user enough to give personalized advice (like a friend would) | 5 | 10 |
|   AI does not register new information when reprompted | 1 | 1 |
|   AI lacking visual modality (diagram, video) to illustrate exercise, yoga, etc | 1 | 1 |
|   AI missing recent information (eg current events) | 2 | 2 |
|   AI missing required context | 4 | 5 |
|   Assuming Western context (lacking cultural nuance) | 4 | 6 |
|   Bias and stereotypes | 1 | 1 |
|   Concern that AI knows intimate personal details that the user does not want to share beyond close family and friends | 1 | 1 |
|   Contesting | 0 | 0 |
|     Diet information | 1 | 1 |
|     Medical information | 1 | 1 |
|   Images created are derivative | 1 | 1 |
|   Inaccurate information | 2 | 2 |
|   Information is too generic (or the prompting is not specific enough, design suggestion for AI to ask follow up questions to refine prompt) | 13 | 23 |
|   Information not applicable (too busy for self-care, other constraints) | 2 | 2 |
|   Lack of control in creative outputs (images, etc) | 1 | 1 |
|   Lack of human experience and empathy | 6 | 12 |
|   Lacking credibility due to lack of source information | 1 | 1 |
|   Lacking diversity | 1 | 1 |
|   Lacking nuance | 2 | 3 |
|   Limits existing creative self-care practice | 1 | 1 |
|   Not consistent depending on prompt | 1 | 1 |
|   Privacy and security concerns (use data for training, data leak) | 4 | 5 |
|   Resource created too close to prompt | 2 | 3 |
|   Web-sources not high-quality (perplexity) | 1 | 1 |
| Modalities | 0 | 0 |
|   Chat-based | 19 | 57 |
|   Multi-modal | 12 | 20 |
| Practices | 0 | 0 |
|   Advice Seeking | 21 | 105 |
|     Cultural Wellbeing | 3 | 6 |
|       Culture-specific advice | 3 | 6 |
|     Digital Wellbeing | 3 | 5 |
|       Digital detox, manage digital distractions, limit screen time or social media | 3 | 4 |
|       Start new habit (reading) | 1 | 1 |
|     Emotional Wellbeing | 4 | 6 |
|     Medical Advice | 2 | 2 |
|       Advice on after-care post tooth removal | 1 | 1 |
| Potential diagnosis for medical condition | 1 | 1 |
| Mental Wellbeing | 14 | 23 |
| Coping strategies for feeling overwhelmed | 1 | 1 |
| Mental health self-care suggestions | 3 | 4 |
| Mindfulness and breathing advice | 2 | 2 |
| Post-work self-care ideas | 1 | 1 |
| Prompts for gratitude diary | 1 | 1 |
| Stress advice | 5 | 6 |
| Stress and anxiety advice | 1 | 1 |
| Suggestions for NY resolutions (self-care) | 1 | 2 |
| Suggestions on how to support anxiety around teaching a coding class | 1 | 1 |
| Time management, productivity advice | 4 | 4 |
| Physical Wellbeing | 16 | 35 |
| Diet and cooking advice | 7 | 10 |
| Exercise advice | 7 | 7 |
| Physical health and wellbeing advice | 3 | 4 |
| Pregnancy advice | 2 | 6 |
| Sleep hygiene | 4 | 5 |
| Workout routine | 1 | 2 |
| Yoga instructions | 1 | 1 |
| Social Wellbeing | 12 | 22 |
| Advice for toddler to help deal with babies crying | 1 | 1 |
| Advice on how to approach difficult conversations | 2 | 2 |
| Advice to support friend | 2 | 2 |
| Self-care hackathon ideas | 1 | 1 |
| Social connection | 1 | 1 |
| Social situations, conversations, relationship advice, setting boundaries (with other people) | 4 | 8 |
| Suggestions for leisure activities (with friends) | 4 | 5 |
| Suggestions for self-care related activities to implement within coding club to help improve the wellbeing of members | 1 | 1 |
| Travel advice | 1 | 1 |
| Suggestions for books, music, movies, games, etc | 4 | 6 |
| Recommended songs for a playlist | 3 | 3 |
| Mentorship | 4 | 6 |
| Confidant (to vent, discuss situations about friends) | 1 | 1 |
| Online friend | 1 | 1 |
| Sleep coach | 1 | 1 |
| Therapist | 1 | 2 |
| Triathlon coach | 1 | 1 |
| Resource Creation | 11 | 26 |
| Coloring in | 3 | 4 |
| Create card for family member (personal gift and creativity) | 1 | 2 |
| Guided mediation for daughter | 1 | 1 |
| Guided meditation based on prompts | 5 | 7 |
| Image generation as inspiration for drawing as a self-care activity | 2 | 2 |
| Image of nature to promote mindfulness | 1 | 2 |
| Origami drawing (craft, mindfulness) | 1 | 1 |
| Paint by numbers image | 1 | 1 |
| Prayer plan | 1 | 1 |
| Quiz | 1 | 1 |
| Short story to read | 1 | 1 |
| Sleep schedule | 2 | 2 |
| Thesis writing plan | 1 | 1 |
| Social Simulation | 1 | 2 |
| Testing social situations | 1 | 2 |
| Therapeutic Self-Expression | 9 | 29 |
| Create mood diary with images | 1 | 1 |
| Creating images for self-expression (rather than reflection) | 2 | 6 |
| Creating images to reflect on feelings (emotion regulation) | 5 | 16 |
| Creating music | 1 | 2 |
| Journalling suggestions | 2 | 3 |
| Qualities of Generative AI | 0 | 0 |
| Adaptivity | 0 | 0 |
| Personalization | 12 | 26 |
| Collaboration | 1 | 1 |
| Additional therapy support tool | 3 | 5 |
| Brainstorming | 4 | 4 |
| Co-creating music | 1 | 2 |
| Reducing mental load | 2 | 5 |
| Suggests related questions for reflection and to refine prompts (only Perplexity) | 2 | 4 |
| Companionship | 1 | 2 |
| Enjoyable | 4 | 8 |
| Feels like a human - signs of humanizing anthropomorphizing the AI (media equation, treat like a human) | 6 | 9 |
| Generative power | 0 | 0 |
| Analyzing data | 1 | 3 |
| Creative | 8 | 18 |
| Mimic human communication | 4 | 6 |
| Impartiality | 1 | 2 |
| Does not get tired | 2 | 4 |
| Non-judgmental | 4 | 6 |
| Objective | 2 | 4 |
| Introductory information - shortcut - good to learn or kickstart new interests or habits | 2 | 3 |
| Proactivity | 0 | 0 |
| Initiative | 1 | 1 |
| Relevant | 4 | 7 |
| Information increase diversity of self-care practices | 1 | 1 |
| Less overwhelming compared with Google search | 2 | 3 |
| Refining through prompts | 3 | 3 |
| Relevant information based on previous prompts (history) | 3 | 4 |
| Trustworthy | 3 | 4 |
| Requires checking in certain contexts | 3 | 3 |
| Trust from cross-references to web sources (Perplexity) | 3 | 3 |
| Type of AI | 0 | 0 |
| Adobe Firefly | 1 | 1 |
| Bing | 1 | 1 |
| Bing Image | 9 | 14 |
| ChatGPT | 21 | 65 |
| DALLE-2 | 2 | 2 |
| Guided.rest | 3 | 4 |
| Kai Wellbeing AI | 1 | 1 |
| Microsoft Copilot | 1 | 1 |
| Perplexity | 5 | 11 |
| Replika | 1 | 1 |
| Soundraw | 1 | 2 |
| Zen Neurobit (personalised meditation) | 1 | 1 |